\documentclass[debug,preprint]{epl}

\usepackage{cite}
\usepackage{amsmath}
\usepackage{amssymb}

\shorttitle{Supercurrent in ferromagnetic weak links}
\shortauthor{T. T. Heikkil\"a \etal}

\title{Non-equilibrium supercurrent through mesoscopic ferromagnetic
weak links}
\author{T. T. Heikkil\"a\inst{1,2}\thanks{E-mail:
\email{Tero.T.Heikkila@hut.fi}} \and F. K. Wilhelm\inst{1,3} \and
G. Sch\"on\inst{1}}  

\institute{
        \inst{1}Institut f\"ur Theoretische Festk\"orperphysik --
        Universit\"at Karlsruhe, D-76128 Karlsruhe, Germany\\
        \inst{2}Materials Physics Laboratory -- Helsinki University of
        Technology, FIN-02015 HUT, Finland\\
        \inst{3}Quantum Transport Group -- TU Delft, 2600 GA Delft, The
        Netherlands} 

\pacs{74.50.+r}{Proximity effects, weak links, tunnelling phenomena,
and Josephson effects}
\pacs{71.70.Ej}{Spin-orbit coupling, Zeeman and Stark splitting,
Jahn-Teller effect} 
\pacs{75.30.Et}{Exchange and superexchange interactions}

\begin{document}

\date{\today}

\maketitle
\begin{abstract}
We consider a mesoscopic normal metal, where the spin
degeneracy is lifted by a ferromagnetic exchange field or
Zeeman splitting, coupled to two superconducting reservoirs. 
As a function of the exchange field or the distance between the
reservoirs, the supercurrent through this device
oscillates with an exponentially decreasing envelope. 
This phenomenon is similar to the tuning of a 
supercurrent by a non-equilibrium quasiparticle
distribution between two voltage-biased reservoirs.  
We propose a device combining the exchange field and
non-equilibrium effects, which allows us to observe a range of novel
phenomena. For instance, part of the field-suppressed supercurrent can
be recovered by a voltage between the additional probes. 
\end{abstract}

Externally controlled weak links in mesoscopic superconducting circuits
have been at the focus of interest in recent years \cite{baselmans}. 
The possibility to control the quasiparticle distribution 
by external voltage probes allows tuning the supercurrent through the
device (mesoscopic SNS transistors). 
It has been predicted that devices with tunnel junctions
\cite{volkov} and systems with good metallic contacts 
\cite{wilhelm} can enter a peculiar 
mesoscopic non-equilibrium state at low temperatures, which
even allows reversing the supercurrent, turning the system into a
$\pi$-junction. This phenomenon has been verified  experimentally
\cite{baselmans}.  
 
Another phenomenon of high interest in superconducting mesoscopics is
the combination of ferromagnetic (F) elements with superconductors (S)
\cite{jiang,demler,lazar,fazio,seviour}. A strong exchange 
interaction $h$ in the ferromagnet is expected to suppress the
superconducting proximity effect, and hence also the supercurrent.
(Several recent experiments \cite{petrashov,lawrence,giroud} do not
confirm this expectation, a fact which, at this stage, is not understood.) 
For weak fields, the supercurrent through a SFS weak link and the
transition temperature of a SF multi-layer are predicted to oscillate
\cite{buzdin,radovic,khusainov} as a
function of the field, or of the width $d$ of the ferromagnet.
The latter defines a characteristic energy scale,
the Thouless energy, which in the diffusive limit is $E_{\rm T}=\hbar D/d^2$,
proportional to the diffusion constant $D$.

In this article, we show that the non-equilibrium-controlled
supercurrent in mesoscopic SNS transistors\cite{wilhelm}
 and the supercurrent in 
SFS weak links are formally equivalent, although one is
tuned by varying the distribution function, while the other is
controlled by modifications of equilibrium spectral functions. Combining
the two phenomena, we can recover by an applied voltage part of the
supercurrent which is suppressed by the exchange field. Thereby, one
can measure the exchange field in the weak link. 

\begin{figure}
\includegraphics{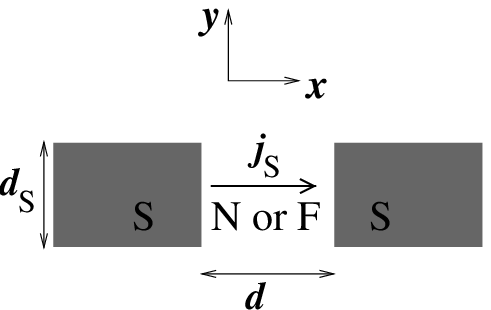}
\includegraphics[clip]{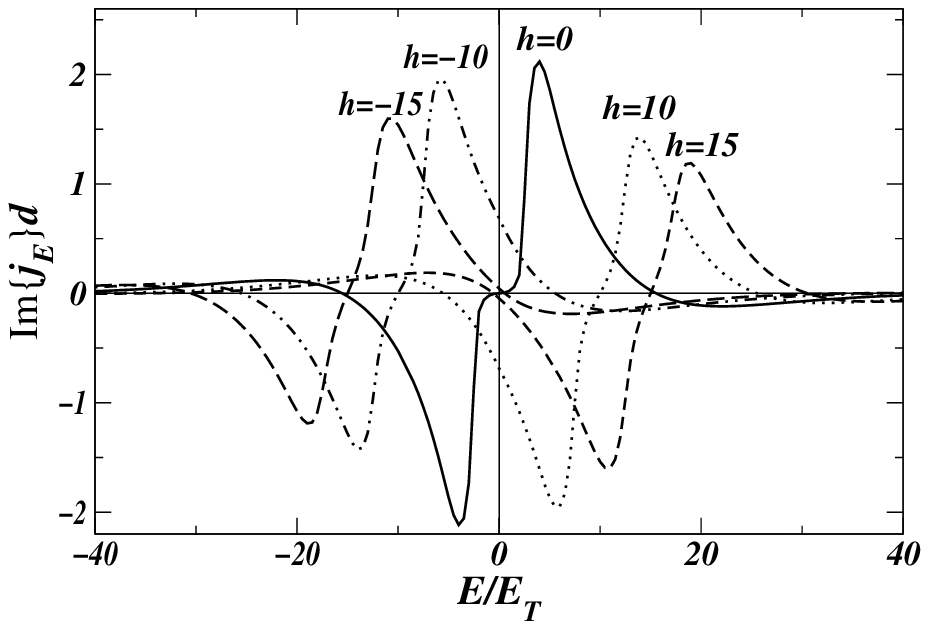}
\caption{Schematic picture of the studied S-F-S structure.}
\label{fig:twoprobe}
\caption{Spectral supercurrent for different exchange fields $h$ at
$\phi=\pi/2$ as a function of energy. The exchange fields are
expressed in the units of the Thouless energy $E_{\rm T}$. The
variation in the peak heights is due to a finite magnitude of the
order parameter $\Delta=50 E_{\rm T}$.} 
\label{fig:spectralsc}
\end{figure}

For definiteness we consider a quasi-one-dimensional system depicted
schematically in fig.~\ref{fig:twoprobe} assuming a three-dimensional
system with structural changes only in one direction. The
magnetism in the weak link, or the Zeeman splitting, is accounted for
by the energy $\sigma h$ of a spatially homogeneous exchange field
coupling to the electron spin $\sigma = \pm 1$. In the diffusive
limit, the system can be described by the Usadel equation for 
quasiclassical Green's functions \cite{demler,fazio,BWBSZ}. While the
equilibrium results of the present work can also be obtained in the
imaginary-time Matsubara formalism,  we have chosen to use the
real-time Keldysh technique in order to include also non-equilibrium
processes. Then we have 
\begin{align}
D\partial_x^2\theta=&-2i(E-\sigma h)\sinh \theta + 2\Delta \cosh
\theta+ \frac{D}{2} (\partial_x \chi)^2 \sinh 2
\theta\label{eq:usadel1}\\
j_E(E,h)&=\sinh^2 \theta \partial_x \chi\; , \qquad \partial_x
j_E=0 \, ,
\label{eq:usadel2}  
\end{align}
where $\theta(E,h,x)$ and $\chi(E,h,x)$ are complex variables
parametrising the quasiclassical diagonal and off-diagonal Green's
function $G(E,h,x) = \cosh(\theta) $ and $F(E,h,x) = \sinh(\theta)
\exp(i\chi)$. For a system of length $d$, eq.~(\ref{eq:usadel1})
introduces a natural energy scale $E_T=D/d^2$. Hence, one way to
tune the relevant energies is by varying the length $d$.
Deep in the superconducting electrodes the exchange field or Zeeman
splitting vanishes. For simplicity, we
assume a bulk BCS solution up to the interfaces,
$\theta_{\rm S}=\arctan(\Delta/E)$, $\chi_{\rm S}=\pm 
\phi/2$ in the superconducting electrodes with amplitudes $\Delta$ and
phase difference $\phi$ of the order parameters of the two
superconductors. Furthermore, we assume clean interfaces, and neglect the
reduction of Andreev reflection expected in spin-polarised systems
\cite{dejong,falko,jedema}. We expect the error due to these
approximations to be only quantitative (for the latter point, see the
discussions below).  

The imaginary part of the conserved `spectral supercurrent', $j_E$, in
eq.~(\ref{eq:usadel2}) enters into the observable supercurrent as
\begin{equation}
j_{\rm S}(\phi)=\frac{d}{4}\sum_{\sigma=\pm 1} g_{{\rm N}\sigma}
\int_{-\infty}^\infty \upd E (1-2f(E)) \text{Im}\{j_E(E,\sigma h)\}. 
\label{eq:observablesc}
\end{equation}
Here $f(E)$ is the distribution function of quasiparticles in the
weak link, which in the absence of applied voltages  reduces to
the equilibrium  Fermi distribution
$f^{\text{eq}}$. Furthermore, $g_{{\rm N}\sigma}=2e^2 N_{0\sigma}
D_\sigma$ is the normal-state conductivity for spin $\sigma$, and
$N_{0\sigma}$ is the corresponding normal-state density of states 
at the Fermi level. Our approach (eqs.~(\ref{eq:usadel1}), 
(\ref{eq:usadel2})) assumes spin-independent densities of states and
diffusion constants. It is valid at low fields $h$,
when the variation in the densities of states is small,
$N_{0\uparrow}-N_{0\downarrow} \ll
(N_{0\uparrow}+N_{0\downarrow})/2$. In this case we may put  
$g_{{\rm N}\uparrow}=g_{{\rm N}\downarrow} \equiv g_{\rm N}$. The
distribution function $f$ in general is obtained from kinetic
equations \cite{BWBSZ}, but for the moment, we assume thermal equilibrium.  

It is instructive to see how the spectral supercurrent Im$\{j_E\}$
depends on energy $E$ and exchange fields $h$. It is  
plotted in fig.~\ref{fig:spectralsc} for a phase difference
$\phi=\pi/2$ between the superconducting electrodes. For $h=0$, the
function Im$\{j_E\}$ is antisymmetric around the Fermi
surface. At low energies $E \lesssim E_{\rm T}$, it vanishes until some
phase-dependent $E_c(\phi)$. At larger energies it
increases sharply, and then decreases exponentially, oscillating
between positive and negative values. The exchange field 
shifts the position of the symmetry point from $E=0$ to $E=\sigma
h$ and for a superconducting gap $\Delta$ of the order of $h$,
distorts the symmetry. Since  $\Delta$ serves as an upper cutoff,
which is not shifted, the overall magnitude of the spectral
supercurrent decreases when $h$ becomes comparable to $\Delta$.  

\begin{figure}[t]
\twofigures[clip,width=.4375\linewidth]{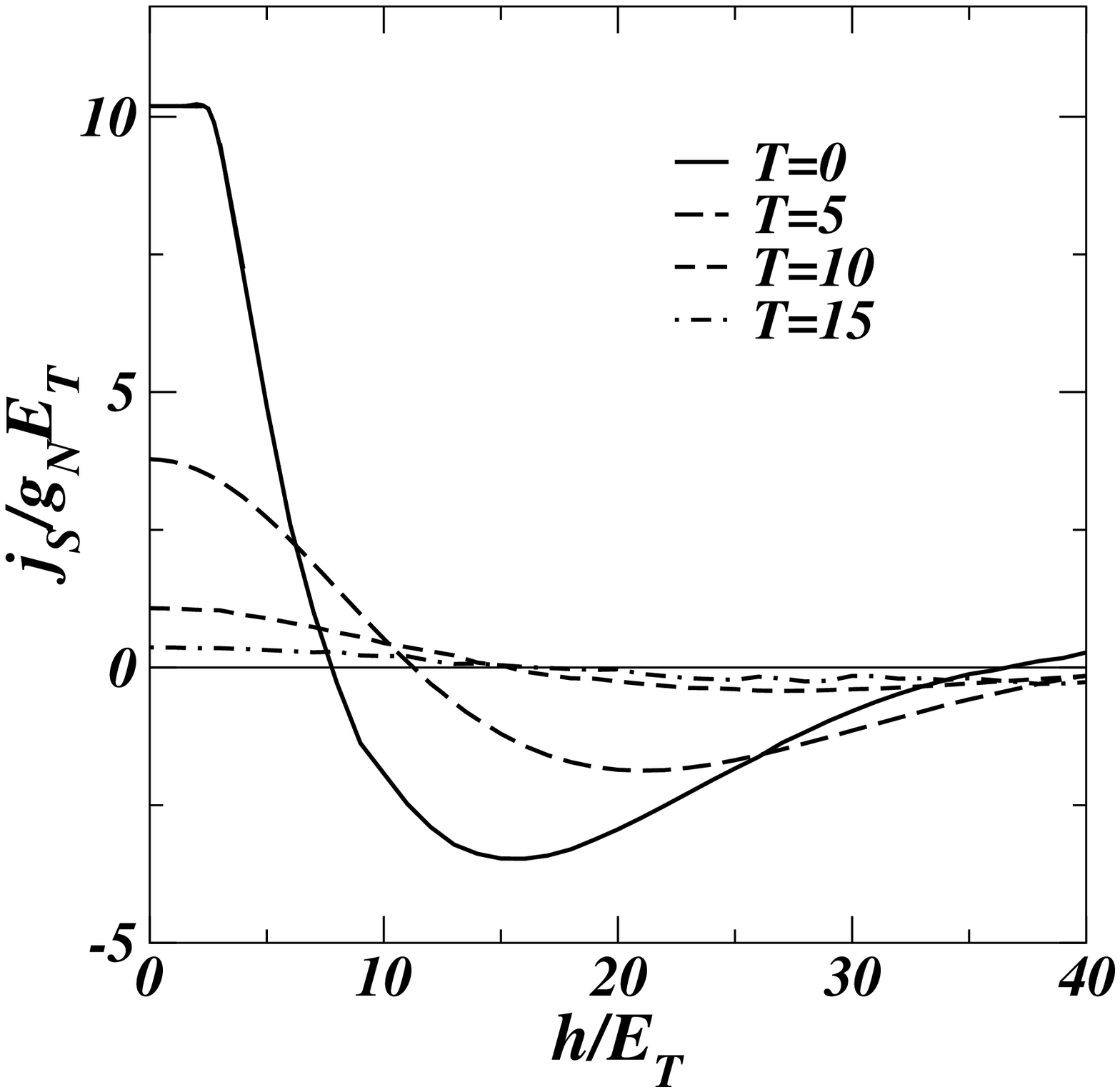}{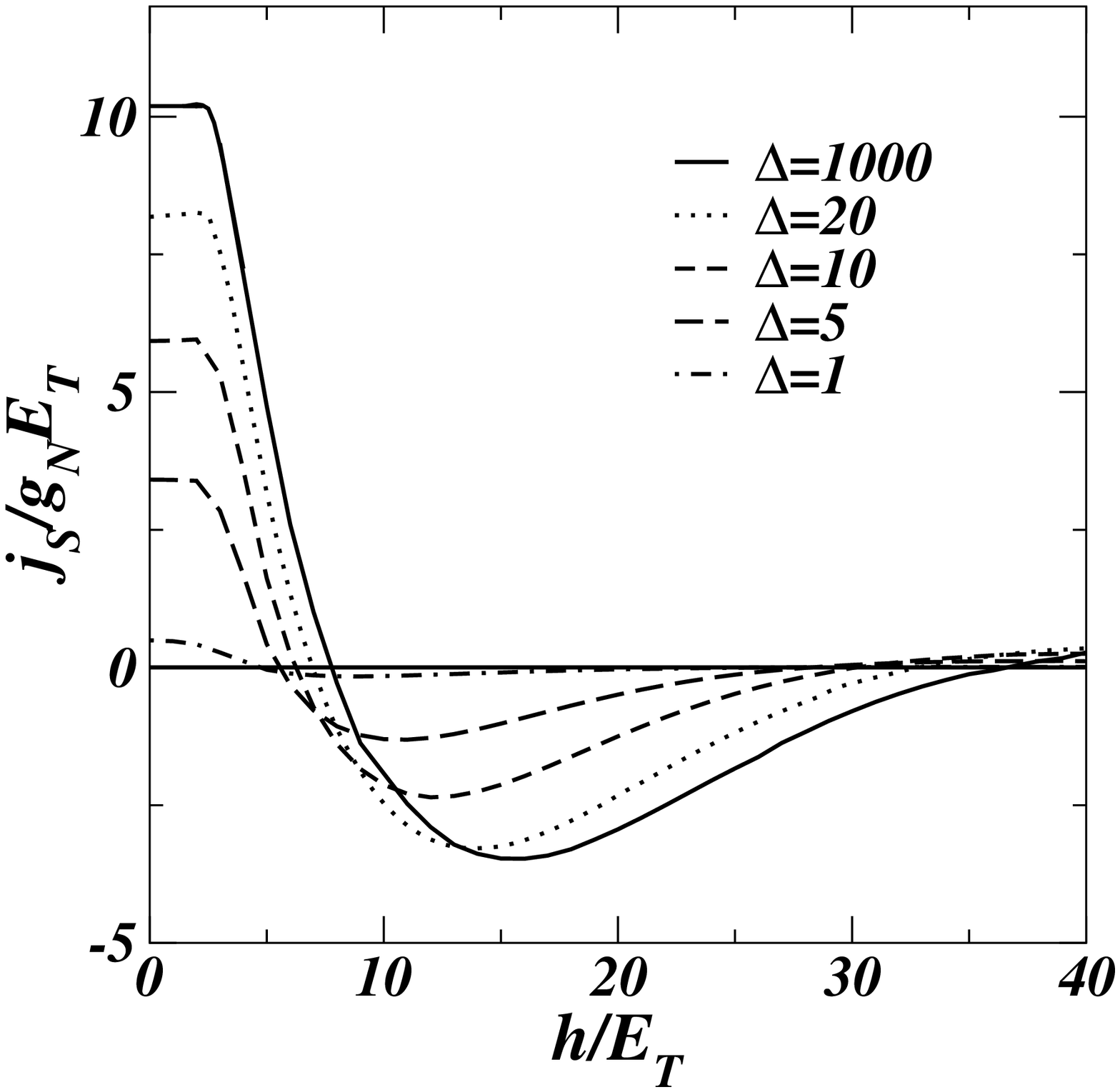}
\caption{Supercurrent $j_{\rm S}(\phi=\pi/2)$ as a function of exchange
field $h/E_{\rm T}$ through the structure depicted in
fig.~\ref{fig:twoprobe} for different temperatures $T/E_T$. The
superconducting order parameter $\Delta=1000 E_T$.} 
\label{fig:IcvshwithT}
\caption{SFS supercurrent $j_{\rm S}(\phi=\pi/2)$ as a function of
exchange field $h/E_{\rm T}$ for different values $\Delta$ of the
superconducting order parameter at $T=0$.} 
\label{fig:IcvshwithDelta}
\end{figure}

In equilibrium we have $1-2f(E)=\tanh(E/2T)$. This term and the sum of
the spectral supercurrent  $\sum_\sigma j_E(E,\sigma h)$ are antisymmetric around
$E=0$. Hence, for the discussion of the total supercurrent $j_{\rm S}$  we
can concentrate on the part $E > 0$. At low $T
\lesssim E_T$ the supercurrent $j_{\rm S}$ is given by an alternating  
sum over the decreasing areas under the oscillating function
Im$\{j_E\}$ measured from the $E$-axis (see
fig.~\ref{fig:spectralsc}). At $h=0$, the 
positive first term dominates the sum and yields a large supercurrent
$j_{\rm S}$. Increasing $h$ shifts the negative peak from $E<0$ to $E>0$, 
hence decreasing $j_{\rm S}$, and even reversing its sign. At finite
temperature, the low-energy part, $E \lesssim T$, is effectively
cut off, hence  $j_{\rm S}$ decreases in amplitude. This result is
illustrated in   
Figs.~\ref{fig:IcvshwithT} and \ref{fig:IcvshwithDelta}, where
$j_{\rm S}(\phi=\pi/2)$ is plotted as a function of different exchange
fields at different temperatures and for different bulk order
parameters $\Delta$. Analogous results can be obtained for a constant
exchange field by varying the distance $d$ of the superconducting
reservoirs and through it the Thouless energy $E_T$.

\begin{figure}
\begin{center}
\twofigures[clip,width=.4375\linewidth]{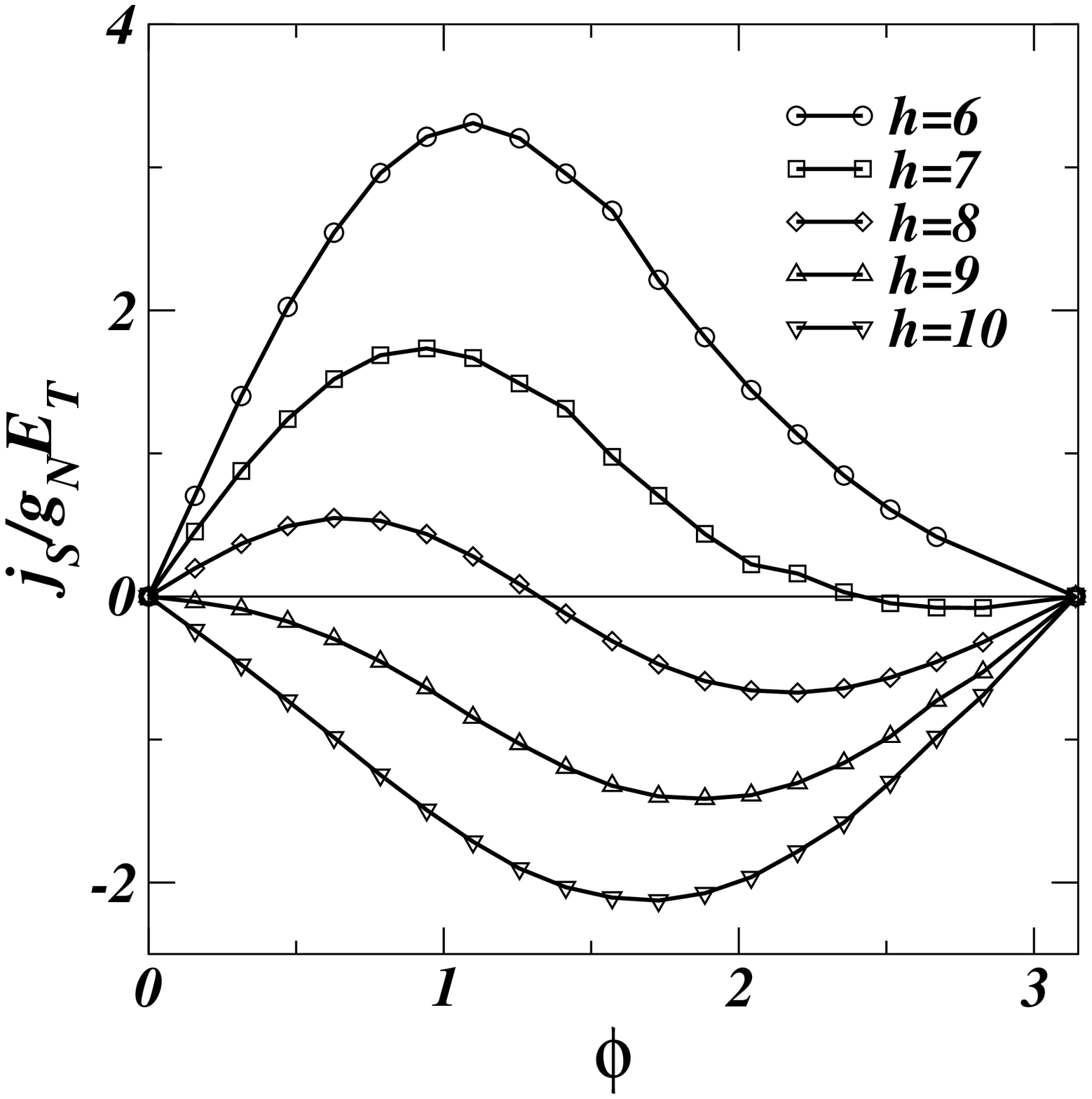}{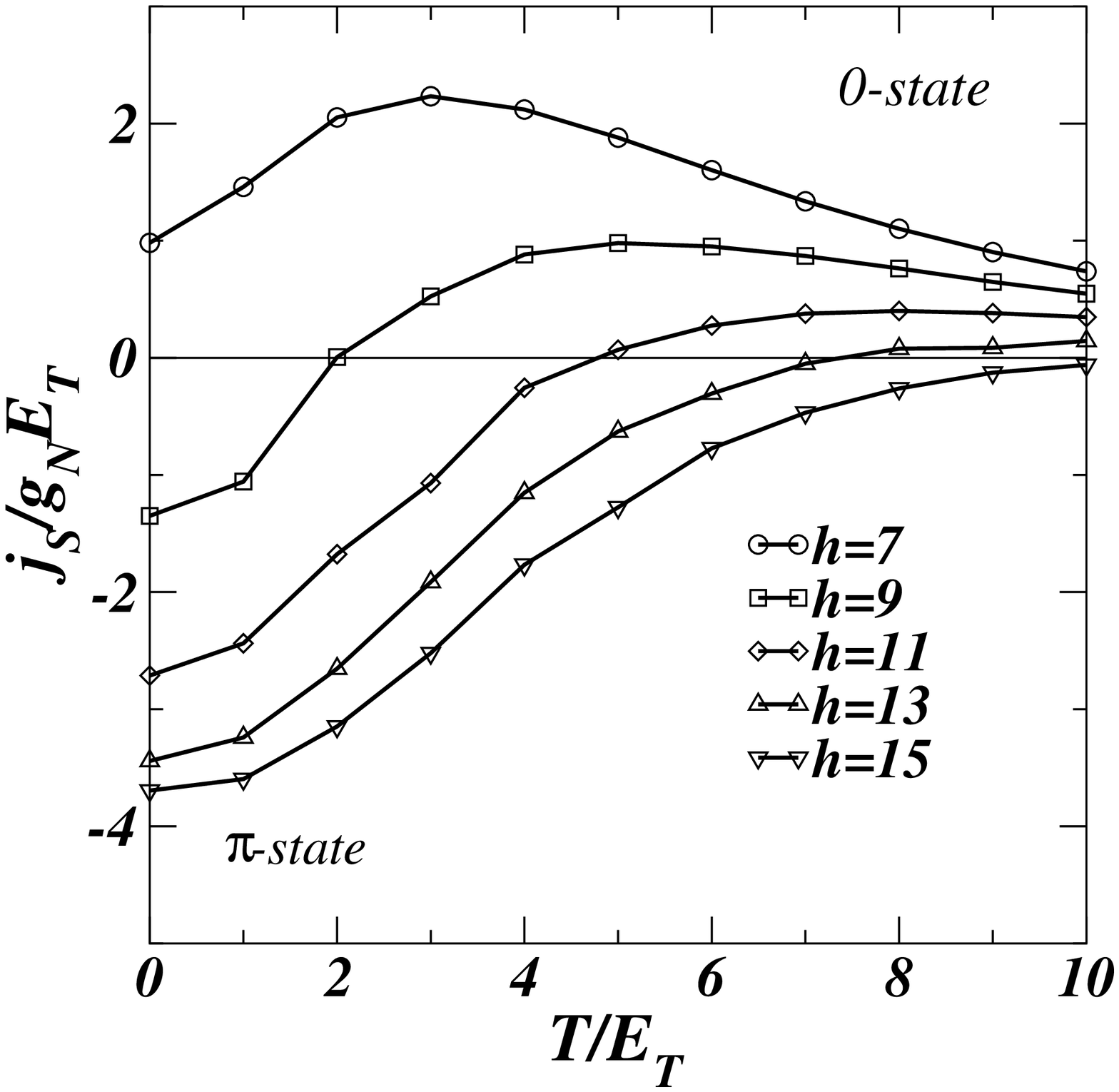}
\caption{Supercurrent $j_{\rm S}(\phi)$ as a function of phase for
different exchange fields $h/E_{\rm T}$ in the regime where the
crossover from the ordinary 0-state to the $\pi$-state occurs for the
first time. Here, $T=0$ and $\Delta=1000E_T$.}
\label{fig:scvsphase}
\caption{Crossover from the $\pi$-state ($j_{\rm S}(\phi=\pi/2)<0$) to
the 0-state ($j_{\rm S}(\phi=\pi/2)>0$) as a function of temperature $T$ for
a few values of the exchange field $h/E_{\rm T}$.} 
\label{fig:crossovervsT}
\end{center}
\end{figure}

In the regime where $j_{\rm S}(\phi=\pi/2)$ is negative, the
junction forms a so-called `$\pi$-junction' \cite{buzdin}, since
the ground state of the system, with no supercurrent
flowing between the two superconductors, is reached for a phase
difference equal to $\pi$. The supercurrent--phase relation for
different exchange  
fields is plotted in fig.~\ref{fig:scvsphase}, showing the crossover
from an ordinary behaviour to a $\pi$-state. A closer analysis of
fig.~\ref{fig:IcvshwithT} shows that the precise value of $h/E_{\rm
T}$ where the crossover occurs depends weakly on temperature,
since higher values of $T$ smoothen the oscillations
between positive and negative contributions to $j_{\rm S}$. At $h=0$,
$j_{\rm S}(\phi=\pi/2)$ is positive for any $T$. It remains positive
as long as the thermal energy dominates over the exchange, $T\gg h$.
With increasing $T$ the cross-over to a $\pi-$junction is shifted
towards higher fields. This dependence was probably observed in
ref.~\cite{veretennikov}. It is an alternative way to verify
the current reversal to what has been discussed in previous  
proposals, where typically one requires many different samples with
varying widths~\cite{jiang} but otherwise equal parameters. 
The crossover is illustrated in fig.~\ref{fig:crossovervsT} for a few
values of $h/E_T$.

\begin{figure}[ht]
\twofigures[clip,width=.4375\linewidth]{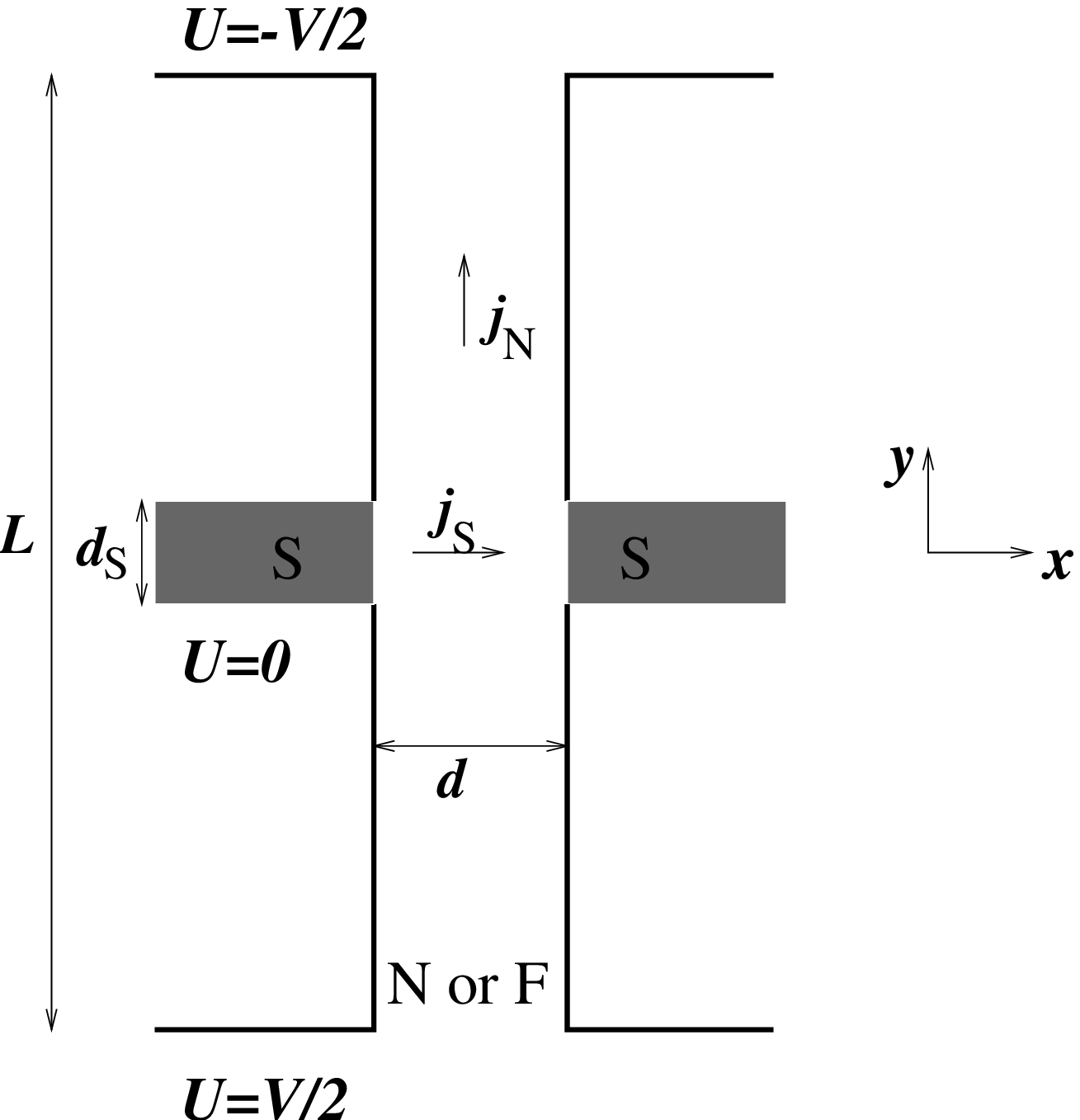}{scvsVandhtwo.eps}
\caption{Four-probe setup for studying the non-equilibrium effects on
the supercurrent. It is assumed that $L \gg d_{\rm S}$ and that the
superconductors lie in the middle of the normal wire ($y=0$) so that the
distribution function has the two-step form between the
superconductors. Furthermore, we expect that the four-probe setup does
not notably alter the spectral supercurrent obtained from a
quasi-one-dimensional calculation (for a more detailed discussion, see
refs. \cite{wilhelm,wilhelmpro}).} 
\label{fig:fourprobe}
\caption{Supercurrent $j_{\rm S}(\phi=\pi/2)$ of the four-probe structure at
different fields as a function of the voltage $V$ between the normal
probes. In the calculations for the main picture, the magnitude of the order
parameter was set to $\Delta=100E_{\rm T}$, and at the inset,
$\Delta=10E_{\rm T}$, thereby showing that even when $\Delta$ is of
the order of $h$ and $eV$, a local maximum is obtained at $eV=2h$.}  
\label{fig:voltageandfield}
\end{figure}

By shifting the variable of integration $E$  in
eq.~(\ref{eq:observablesc}) by  $\sigma h$ one finds 
\begin{equation}
j_{\rm S}(\phi)=\frac{dg_{\rm N}}{2}\sum_\sigma \int_{-\infty}^\infty
\upd E (1-f^{\text{eq}}(E+\sigma h)) {\text{Im}}\{j_E(E)\}.
\label{eq:equivalence}
\end{equation}
The shifted distribution function $f=1/2\sum_\sigma f^{\rm eq}(E+\sigma h)$
has the same form as the two-step distribution function measured in
ref.~\cite{pothier}. There it appeared as the solution of a kinetic
equation  in the centre of a diffusive metal between two normal probes with
voltage $eV=\pm 2h$ in the limit where the inelastic scattering length
is longer than the distance between the two normal reservoirs. 
The spectral supercurrent in general still depends on the exchange field 
via the boundary conditions. However, if the superconducting gap
$\Delta$ is much larger than the exchange field, $\Delta \gg h$,
this dependence can be neglected. In this limit, the supercurrent in
the presence of an exchange field is the same as for 
a non-equilibrium distribution four-probe structure  described in
ref.~\cite{wilhelm} (see fig.~\ref{fig:fourprobe} for a schematic picture). 

It is interesting to note that this behaviour of the diffusive-limit
supercurrent as functions of the exchange field and the external
potential is very similar to the supercurrent through a multi-probe
structure in the clean limit. This limit has been described by
Dobrosavljevi$\acute{\text{c}}$-Gruji$\acute{\text{c}}$ \etal
\cite{dobro} for the ferromagnetic two-probe case and by van Wees
\etal \cite{vanwees} including a voltage in a non-magnetic three-probe
setup. In this case, the supercurrent is carried by the Andreev
levels, whose energies are controlled by the exchange field \cite{dobro}, and
whose occupation can be tuned by the voltage \cite{vanwees}. With both
parameters, for example, the system can be driven into a $\pi$-state.

We can combine the effects of exchange field and non-equilibrium
distribution~\cite{wilhelm,wilhelmpro} by considering the structure in
fig.~\ref{fig:fourprobe}, where the magnetic material is placed between 
superconductors and normal voltage leads. 
Here, the distribution function is 
\begin{equation}
f(E,y)=(\frac{1}{2}-\frac{y}{L})f^{\text{eq}}(E+eV/2) + (\frac{1}{2} +
\frac{y}{L})f^{\text{eq}}(E-eV/2),
\end{equation}
exhibiting the two-step form observed by Pothier et
al. \cite{pothier}. In this case, if the 
superconducting reservoirs are located around $y=0$ and provided 
$\Delta \gg h,eV$, the observable supercurrent can be written as a sum
of four terms,
\begin{equation}
\begin{split}
j_{\rm S}(\phi)=\frac{dg_{\rm N}}{8} \int_{-\infty}^\infty \upd E
(1-2f^{\text{eq}}(E)) {\text{Im}}\{&j_E(E-h-eV/2)+
j_E(E-h+eV/2)+\\&j_E(E+h-eV/2)+j_E(E+h+eV/2)\}. 
\end{split}
\end{equation}
For example, if the potential is exactly twice the exchange field,
$eV=2h$, due to the antisymmetry of Im$\{j_E\}$, we have
\begin{equation}
j_{\rm S}(\phi)=\frac{1}{2}\left(j_{\rm S}^{\text{SFS}}(\phi,0)+
j_{\rm S}^{\text{SFS}}(\phi,2h)\right) \approx
\frac{1}{2}j_{\rm S}^{\text{SFS}}(\phi,0).  
\label{eq:screcover}
\end{equation}
The latter approximate equality holds if $h \gg E_{\rm T}$. Here,
$j_{\rm S}^{\text{SFS}}(\phi,h)$ is the supercurrent through the SFS
structure with the exchange field $h$ in the weak link. Hence, one
can use the external potential to recover half of the zero-field
supercurrent. This is illustrated in fig.~\ref{fig:voltageandfield},
where the supercurrent of the four-probe structure is plotted as a
function of voltage with a few magnitudes of fields. 

The results summarised by eq.~(\ref{eq:screcover}) provide a way
to measure the exchange field and at the same time to
explore the applicability of the simplified model
for the ferromagnet used here and previously
\cite{buzdin,fazio,seviour}.  When the voltage-dependent supercurrent
$j_{\rm S}(V)$ reaches a 
maximum, $eV$ should equal $2h$. Deviations could occur as, for
instance, this model neglects the band-structure effects \cite{mazin} 
important in the ferromagnets. Also, to be able to measure the actual
supercurrent through a typical ferromagnet with Curie temperature
$T_{\text{Cu}} \gg \Delta$, the  ratio $h/E_T$ has to be made small by
fabricating very thin weak links. Moreover, our assumption of the
diffusive regime requires $d \gg l_{\text{el}}$, and a quantitative
agreement cannot be expected for thin structures. Finally, due to
the strong 
electron-electron interactions in ferromagnets, producing a short
inelastic relaxation length, the normal probes should be fabricated rather
close to each other to obtain the two-step form for the distribution
function. For  conventional ferromagnets the exchange field is large,
which would correspond to enormous voltages.
However, we expect that our model is approximately valid for
setups constructed from ferromagnetic alloys with $T_{\text{Cu}}$ of
the order of the superconducting critical temperature
\cite{veretennikov}, or in situations where $h$ can be related to the
Zeeman splitting in magnetic fields much weaker than the
superconducting critical field.  

In summary, we have calculated the supercurrent through a
ferromagnetic weak link as a function of the exchange field in the
ferromagnet. In the calculations, the Keldysh technique was used to
provide a description of non-equilibrium effects. We found that when
$\Delta \gg h$, the problem is formally equivalent to the four-probe
measurement of the supercurrent through a normal-metal weak
link. Furthermore, we showed that applying a non-equilibrium potential
in the transverse direction, one can recover half of the supercurrent
of a ferromagnet with an exchange field $h \gg E_{\rm T}$, as compared
to the supercurrent in the absence of $h$. 

\acknowledgments

We thank B. Pannetier, M. Giroud and M. Fogelstr\"om for
discussions. This work was supported by the Helsinki University of
Technology, the DFG through SFB 195 and the EU through the EU-TMR
network ``Superconducting nano-circuits''. While writing this paper, 
we became aware of a related work by Yip \cite{yip}, who introduced
the exchange-field term as the Zeeman splitting of the
current-carrying density of states.


\begin{thebibliography}{25}

\bibitem{baselmans} \Name{Morpurgo A., van Wees B. J. \and Klapwijk T. M.} 
\REVIEW{Appl. Phys. Lett.}{72}{1998}{966}; \Name{Baselmans J. J. A.,
Morpurgo A., van Wees B. J., \and Klapwijk
T. M.} \REVIEW{Nature}{397}{1999}{43}.  
\bibitem{volkov} \Name{Volkov A. F.}
\REVIEW{Phys. Rev. Lett.}{74}{1995}{4730}. 
\bibitem{wilhelm} \Name{Wilhelm F. K., Sch\"on G., \and Zaikin A. D.} 
\REVIEW{Phys. Rev. Lett.}{81}{1998}{1682}.
\bibitem{jiang} \Name{Jiang J. S., Davidovi$\acute{\text{c}}$ D., Reich
D. H. \and Chien C. L.} \REVIEW{Phys. Rev. Lett.}{74}{1995}{314}
\bibitem{demler} \Name{Demler E. A., Arnold G. B. \and Beasley M. R.}
\REVIEW{Phys. Rev. B.}{55}{1997}{15174}.
\bibitem{lazar} \Name{Lazar L., Westerholt K., Zabel H., Tagirov
L. R., Goryunov Yu. V., Garif'yanov N. N. \and Garifullin I. A.}
\REVIEW{Phys. Rev. B}{61}{2000}{3711}.
\bibitem{fazio}\Name{Fazio R. \and Lucheroni C.}
\REVIEW{Europhys. Lett.}{45}{1999}{707}. 
\bibitem{seviour} \Name{Seviour R., Lambert C. J. \and Volkov A. F.}
\REVIEW{Phys. Rev. B}{59}{1999}{6031}.
\bibitem{petrashov} \Name{Petrashov V. T., Antonov V. N., Maksimov
S. V., Shaikhaidarov R. Sh.} \REVIEW{JETP Lett.}{59}{1994}{551};
\Name{Petrashov V. T., Sosnin I. A., Cox I., Parsons A. \and Troadec
C.} \REVIEW{Phys. Rev. Lett.}{83}{1999}{3281}.
\bibitem{lawrence} \Name{Lawrence M. D. \and Giordano N}
\REVIEW{J. Phys.: Condens. Matter}{8}{1996}{L563}.
\bibitem{giroud} \Name{Giroud M., Courtois H., Hasselbach K., Mailly
D. \and Pannetier B.} \REVIEW{Phys. Rev. B}{58}{1998}{R11872}.
\bibitem{buzdin} \Name{Buzdin A. I., Bulaevskii L. N. \and Panyukov
S. V.} \REVIEW{JETP Lett.}{35}{1982}{178}; \Name{Buzdin A. I. \and
Kupriyanov M. Yu.} \REVIEW{JETP
Lett.}{52}{1990}{487}; \SAME{53}{1991}{321}; \Name{Buzdin A. I.,
Bujicic, B. \and Kupriyanov M. Yu}
\REVIEW{Sov. Phys. JETP}{74}{1992}{124}.  
\bibitem{radovic} \Name{Radovi$\acute{\text{c}}$ Z.,
Dobrosavljevi$\acute{\text{c}}$-Gruji$\acute{\text{c}}$ L., Buzdin 
A. I. \and Clem J. R.} \REVIEW{Phys. Rev. B}{38}{1988}{2388};
\Name{Radovi$\acute{\text{c}}$ Z., Ledvij M.,
Dobrosavljevi$\acute{\text{c}}$-Gruji$\acute{\text{c}}$ L., Buzdin
A. I. \and Clem J. R.} \REVIEW{Phys. Rev. B}{44}{1991}{759}.
\bibitem{khusainov} \Name{Khusainov M. G. \and Proshin Yu. N.}
\REVIEW{Phys. Rev. B}{56}{1997}{R14 283}.
\bibitem{BWBSZ} \Name{Belzig W., Wilhelm F. K., Bruder C., Sch\"on
G. \and Zaikin A. D.} \REVIEW{Superlatt. and
Microstruct.}{25}{1999}{1251}.
\bibitem{dejong}\Name{de Jong, M. J. M. \and Beenakker, C. W. J.}
\REVIEW{Phys. Rev. Lett.}{74}{1995}{1657}.
\bibitem{falko}\Name{Fal'ko, V. I., Lambert, C. J. \and Volkov,
A. F.} \REVIEW{JETP Lett.}{69}{1999}{532};
\REVIEW{Phys. Rev. B}{60}{1999}{15394}. 
\bibitem{jedema}\Name{Jedema, F. J., van Wees B. J., Hoving B. H.,
Filip A. T. \and Klapwijk T. M.} [cond-mat/9901323].
\bibitem{veretennikov} \Name{Veretennikov A. V., Ryazanov V. V.,
Oboznov V. A., Rusanov, A. Yu., Larkin, V. A. \and Aarts J.}
\REVIEW{Physica B}{284-288}{2000}{495}; \Name{Aarts J.} private communication.
\bibitem{pothier} \Name{Pothier H., Gu$\acute{\text e}$ron S., Birge
N. O., Esteve D. \and Devoret M. H.}
\REVIEW{Phys. Rev. Lett.}{79}{1997}{3490}. 
\bibitem{dobro}
\Name{Dobrosavljevi$\acute{\text{c}}$-Gruji$\acute{\text{c}}$ L.,
Ziki$\acute{\text{c}}$ R. \and Radovi$\acute{\text{c}}$ Z.}
[cond-mat/9911339]. 
\bibitem{vanwees} \Name{van Wees B. J., Lenssen K.-M. H. \and Harmans
C. J. P. M.} \REVIEW{Phys. Rev. B}{44}{1991}{470}.
\bibitem{wilhelmpro} \Name{Wilhelm F. K., Sch\"on G. \and Zaikin
A. D.} \REVIEW{Physica B}{280}{2000}{418}.
\bibitem{mazin} \Name{Mazin I. I.}
\REVIEW{Phys. Rev. Lett.}{83}{1999}{1427}.
\bibitem{yip} \Name{Yip S. K.} [cond-mat/0002395].
\end{thebibliography}
\end{document}